\documentclass{article}

\usepackage{arxiv}

\usepackage[utf8]{inputenc} 
\usepackage[T1]{fontenc}    
\usepackage{hyperref}       
\usepackage{url}            
\usepackage{booktabs}       
\usepackage{amsmath,amssymb,amsfonts} 
\usepackage{nicefrac}       
\usepackage{microtype}      
\usepackage{graphicx}
\usepackage{algorithm,algorithmic}
\usepackage{pdflscape}
\usepackage{multirow}
\usepackage{array}
\usepackage{authblk}        
\usepackage{xcolor}         
\usepackage{stfloats}       
\usepackage{float}          
\usepackage{multicol}       
\usepackage{cuted}

\newcommand{\bigfirstletter}[2]{{\large\textbf{#1}}#2}

\newcommand{\centeredkeywords}[1]{
  \par\vskip 0.5ex
  \begin{center}
    \small\textbf{Keywords}: #1
  \end{center}
}

\title{Biological Multi-Layer and Single Cell Network-based Multiomics Models – a review}

\author[1,*]{Marcello Barylli}
\author[2,*]{Joyaditya Saha}
\author[3]{Tineke E. Buffart}
\author[3]{Jan Koster}
\author[3,4]{Kristiaan J. Lenos}
\author[3,4]{Louis Vermeulen}
\author[1]{Vivek M. Sheraton}
\affil[1]{Computational Science Lab, Informatics Institute, University of Amsterdam, Amsterdam, The Netherlands}
\affil[2]{Laboratory for Experimental Oncology and Radiobiology (LEXOR), Center for Experimental and Molecular Medicine (CEMM), Cancer Center Amsterdam (CCA), Amsterdam Gastroenterology Endocrinology and Metabolism (AGEM), Amsterdam University Medical Centers (location AMC), Amsterdam, The Netherlands}
\affil[3]{Cancer Center Amsterdam (CCA), Amsterdam University Medical Centers (location AMC), Amsterdam, The Netherlands}
\affil[4]{Oncode Institute, Utrecht, The Netherlands}
\affil[*]{These authors contributed equally to this work}

\begin{document}

\onecolumn
\maketitle

\begin{abstract}
Recent advances in single cell sequencing and multi-omics techniques have significantly improved our understanding of biological phenomena and our capacity to model them. Despite combined capture of data modalities showing similar progress, notably single cell transcriptomics and proteomics, simultaneous multi-omics level probing still remains challenging. As an alternative to combined capture of biological data, in this review, we explore current and upcoming methods for post-hoc network inference and integration with an emphasis on single cell transcriptomics and proteomics. By examining various approaches, from probabilistic models to graph-based algorithms, we outline the challenges and potential strategies for effectively combining biological data types while simultaneously highlighting the importance of model validation. With this review, we aim to inform readers of the breadth of tools currently available for the purpose-specific generation of heterogeneous multi-layer networks.
\end{abstract}

\centeredkeywords{Heterogeneous networks $\cdot$ multi-omics $\cdot$ single cell sequencing}
\vspace{0.5cm}

\begin{multicols}{2}
\raggedcolumns

\section{Introduction}
\label{sec:introduction}
\bigfirstletter{I}{dentifying} the unique characteristics of specific cell populations is key to understanding disease onset and progression \cite{vandesande2023}. A lot of our current understanding of biological phenomena can be attributed to techniques which investigate changes within cells or tissues at different molecular levels (genomic, transcriptomic, proteomic or metabolomic). Regardless of the insights these techniques have provided, inferring cellular heterogeneity from biological data obtained from tissues has remained challenging. Specific deconvolution algorithms have been developed to infer cellular populations from such data based on \textit{a priori} information, which makes them inherently biased and does not permit identification of novel populations of cells \cite{liBulkSinglecellSpatial2021}\cite{kulkarniBulkReviewSingle2019}.

This limitation was addressed with the development of single cell RNA-sequencing (scRNA-Seq). ScRNA-Seq was introduced by Tang et al. in 2009 and has since then revolutionised fundamental \& translational biological research \cite{tangMRNASeqWholetranscriptomeAnalysis2009}. The general workflow of scRNA-Seq critically involves the isolation of single cells from a heterogeneous population, allowing the observation of cell-specific gene expression changes. Since its introduction, further developments in the field of scRNA-Seq have permitted the technique to be scaled-up, increasing throughput and frequency at which biological data is generated and analysed. In conjunction with these improvements, advancements in computational analysis techniques have expanded the types and complexity of analyses performed using the generated data \cite{vandesande2023}\cite{brockley2023}.

Despite the progress made in the field of single cell transcriptomics, a crucial remaining limitation is the fact that single cell transcriptomics only takes into account transcriptomic changes between cells. As such, single cell transcriptomics cannot give a complete picture of the biology underscoring cells. To get a more comprehensive understanding, one can to look to the emerging field of multi-omics, wherein the relationships between different molecular levels are taken into account to explain phenotypic features. In this review, we will discuss the various types of models that can be used to generate such multi-omics networks, the advantages and disadvantages of each, the challenges that lie ahead and how they can each uniquely contribute to a better and more comprehensive understanding of the biology underscoring homeostasis and pathology.

\subsection{Current Challenges with Network Approaches for Multi-Omics Integration}
While significant progress has been made in recent years in developing network-based methods for single-cell analysis and multi-omics integration, alignment of omics data at a single-cell level remains challenging. Single cell sequencing is known to suffer from zero-inflation, corresponding to a high number of missed counts in the reads data\cite{tranSinglecellRNASequencing2021}\cite{wangScGNNNovelGraph2021}, as well as increas\textbf{}ed levels of noise when compared to bulk methods\cite{leou-yangInferringCancerCommon}. Furthermore, incorporation of multiple data sources leads to an increased risk of overfitting, which therefore requires appropriate regularisation\cite{argelaguetComputationalPrinciplesChallenges2021}. Overfitting occurs when a model is fit to each data point, thereby also capturing noise and resulting in poor generalisability. This can be mitigated by, for example, lasso regression. Lasso regression penalises overly complex parameter sets and encourages simpler models via sparsification, setting many of the parameter values to zero. Finally, increasing the resolution and types of data sources places additional demand on computational resources, which must be met through more efficient computational techniques.

While techniques pioneered by 10X Genomics allow relatively robust capture of scRNA information, corresponding cell-specific proteomics and metabolomics data are typically lost. Thus, heterogeneous multilayer network architectures that combine multiple layers of matched biological information are currently underrepresented in the literature. Currently established routine single cell multi-omics includes single cell Assay for Transposase-Accessible Chromatin (scATAC-Seq) and the more recent Cellular Indexing of Transcriptomes and Epitopes (CITE-Seq), which focus on chromatin accessibility and RNA coupled with epitope expression, respectively\cite{buenrostroSinglecellChromatinAccessibility2015}\cite{stoeckius2017}. These methods have proven essential for the inference of gene regulatory networks (GRNs) in particular, providing contextual information for determining causal node relationships \cite{badia-i-mompelGeneRegulatoryNetwork2023}. 

However, explicitly modelling the RNA and protein layers would permit higher model expressivity and interpretability. Recent advances in combined transcriptomics and proteomics captured at a single cell level at high-resolution hold promise for future development of future multi-omics network models \cite{reimegardCombinedApproachSinglecell2021}\cite{katzenelenbogenCoupledScRNASeqIntracellular2020}\cite{gaultCombiningNativeOmics2020}. For example, Intracellular Staining and Sequencing (INs-Seq) enables intracellular protein immunodetection followed by scRNA-seq \cite{katzenelenbogenCoupledScRNASeqIntracellular2020}. While this does not cover the entire proteome, such integrated profiling motivates the construction of focused network models investigating specific biological functions. Another example is given by nanodroplet SPlitting for Linked-multimodal Investigations of Trace Samples (nanoSPLITS),  which robustely profiles $>$ 5000 genes and $>$2000 proteins from a single cell \cite{fulcherParallelMeasurementTranscriptomes2022}. 
While these techniques increase the capacity of single cell combined capture, they have not yet reached equal coverage for all biomolecules. Therefore, single cell multilayer networks for proteomics and RNA at a multi-omics scale currently still requires determining a correspondence between different biomolecules. This is expanded upon in section \ref{sec:anchors}.

\begin{figure}[H]
    \centering
    \includegraphics[width=\columnwidth]{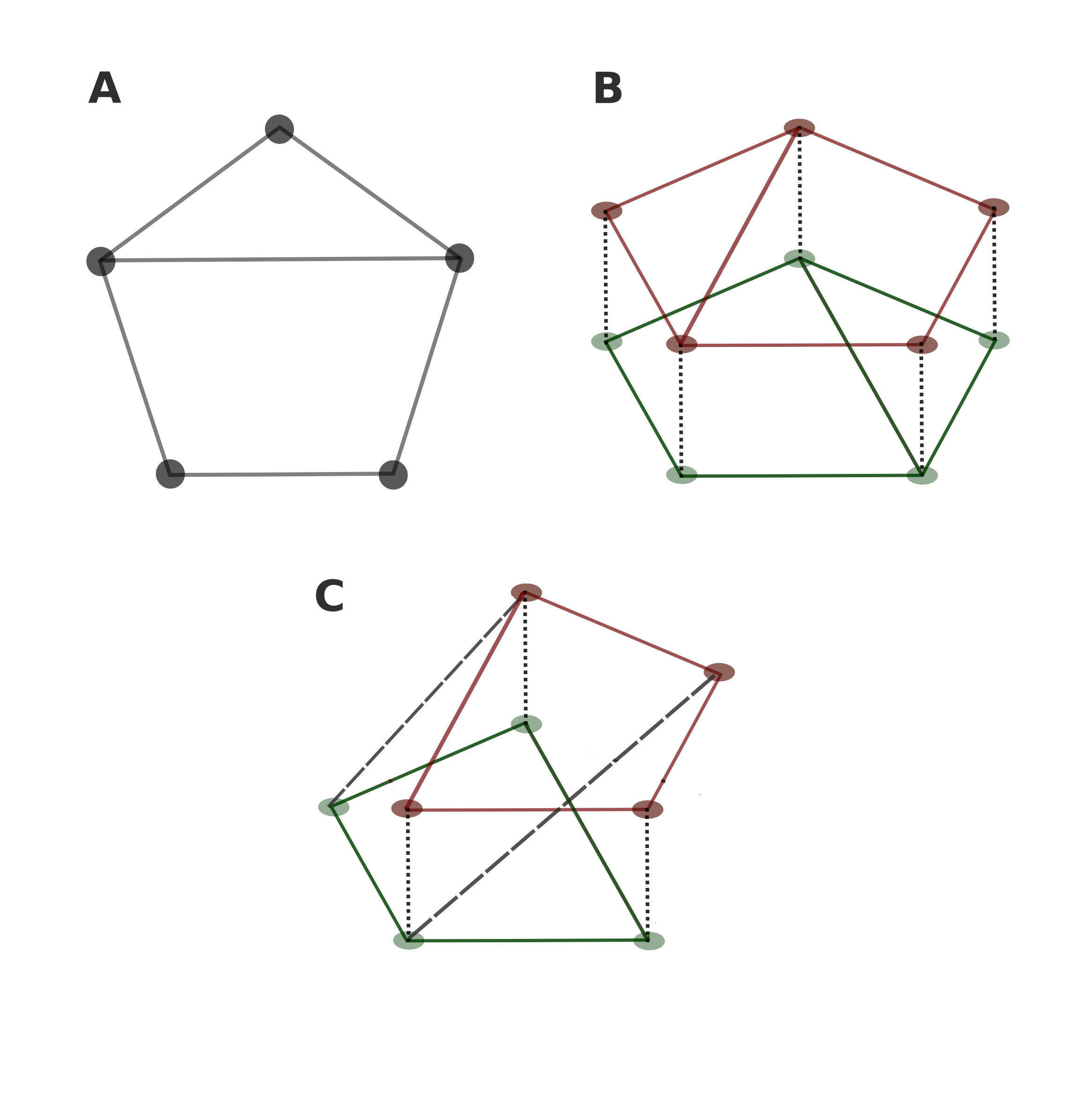}
    \caption[Monoplex, Multiplex and Multi-layer Networks]{\textbf{Network architectures.} (A) A single layer network consisting of a single node and edge set. (B) A multiplex network, where a single no\textbf{}de set is interconnected by multiple edge sets. (C) A more general multilayer network, where node and edge sets do not necessarily overlap, and cross-layer edges are possible between any two nodes of differing layers.}
    \label{fig: network types}
\end{figure}

\subsection{Network Terminology}
Networks, or graphs, are characterised by their non-euclidean structure, encompassing a set of vertices (nodes) interconnected by edges (links) \cite{barabasi2016network}. A graph $G$ is defined as $(G = (V, E))$, where $V$ is the set of nodes and $E \subseteq V \times V$ is the set of edges connecting the nodes\cite{kivelaMultilayerNetworks2014}. Such a network is readily visualised, as seen in Figure \ref{fig: network types}, where \ref{fig: network types}A represents a standard network model. If two nodes are connected by an edge, they are said to be adjacent. $G$ can thus be expressed using an adjacency matrix $A$, the values of which are set to $1$ for connected nodes and $0$ otherwise. Furthermore, edges may be directed or undirected. In the former case, a causal relationship can be established between two entities. 

Multiplex networks, as seen in Figure \ref{fig: network types}B, model a single set of nodes from multiple \textit{aspects} that characterise the nodes in different ways. Each aspect contains a separate edge set. The most flexible model is the heterogeneous  multi-layer network, depicted in Figure \ref{fig: network types}C. These networks may contain multiple, potentially overlapping node sets and edge sets, with cross-layer edges possible between any two nodes of differing layers.

\subsection{Building Networks from Single Cell Data}
\label{sec:building_sc_from_matrices}

Network inference from single cell data begins with a matrix containing cells as columns and genes as rows. Each cell represents a point in high-dimensional gene-space. Techniques such as t-SNE, UMAP, or PCA are commonly used to reduce the dimensionality, upon which cells may be clustered \cite{NIPS2002_6150ccc6}\cite{mcinnesUMAPUniformManifold2020}\cite{jolliffePrincipalComponentAnalysis2016}. Links between cells are established based on their proximity in this reduced space\cite{hetzelGraphRepresentationLearning2021}.

Gene-gene interactions are typically inferred
based on the correlation of their expression values across cells
of a given type. Gene pairs exceeding a predefined similarity threshold form network links. This notion is termed "guilt by association"\cite{cowenNetworkPropagationUniversal2017}. While this omits causal relationships, it serves as a basis for downstream directionality inference.

The result of these two orthogonal views is a cell-type-specific gene network, where each cell type is defined by its internal gene network, and each gene network is shaped by its connection to the gene networks of other cells. This creates a multi-scale "network of networks"\cite{hammoudMultilayerNetworksAspects2020}. To link the gene networks, cell-cell communication information is incorporated using tools like CellPhoneDB\cite{efremovaCellPhoneDBInferringCell2020} or CellChat\cite{jinInferenceAnalysisCellcell2021}, which leverage the ligand-receptor interactions shaping cellular communities. Extracellular metabolites and RNA also perform crucial signalling roles in cellular environments. For example, amino acids may act as signalling molecules that influence cell behavior and gene expression in the tumour microenvironment\cite{eliaMetabolitesTumorMicroenvironment2021}. Extracellular RNAs carried by various particles, including vesicles and lipoproteins, can mediate cell-cell communication through both local and systemic mechanisms\cite{murilloExRNAAtlasAnalysis2019}.

\begin{figure}[H]
    \centering
    \includegraphics[width=\columnwidth]{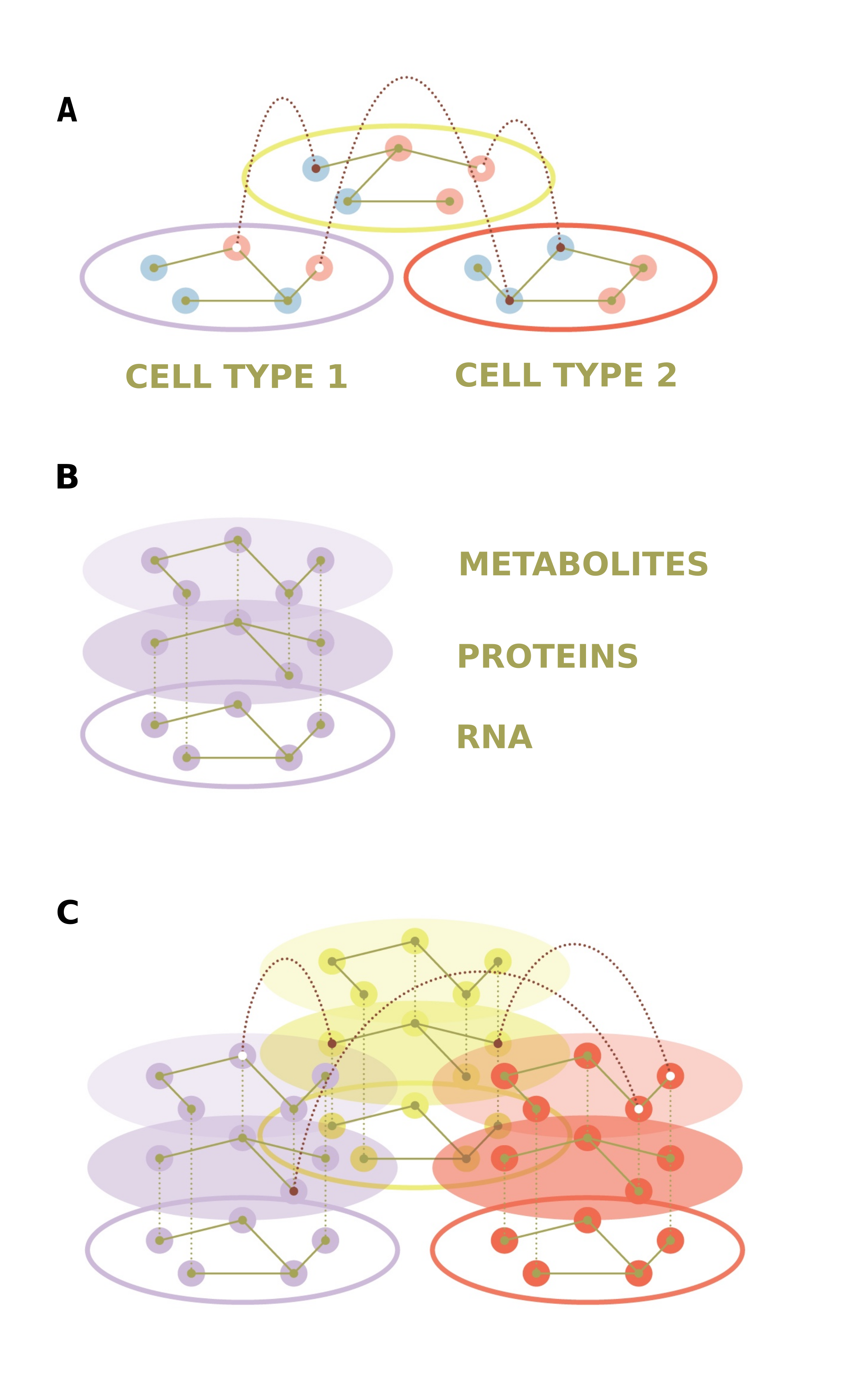}
    \caption{\textbf{Architecture of the heterogeneous multilayer network
of networks.} \textbf{(A)} Network of Networks
\cite{CytoTalkNovoConstruction}.
This represents intra-cell gene regulatory networks, linked to each
other through cell-cell interactions. \textbf{(B)} Heterogeneous
multilayer network model
\cite{leeHeterogeneousMultiLayeredNetwork2020}\textbf{.}
Layers are abstractions of arbitrary modalities. However, they could
also represent relationships described by the central dogma of biology,
with bi-directional interactions. \textbf{(C)} Combined view.
Ligand-receptor interactions are explicitly modelled in the protein and
metabolite layers.}
    \label{fig:multiview_networks}
\end{figure}

Figure \ref{fig:multiview_networks}A represents such a network of networks, constructed from a single level of biological information. Cross-links between cells represent intercellular interactions. Note that this is not a heterogeneous network, as nodes are all of a single data type, eg scRNA. Next to the multi-scale perspective, the multi-layer viewpoint is visualised in Figure \ref{fig:multiview_networks}B. Here, biological relationships corresponding to the translation of transcripts into proteins as well as the corresponding catalytic activity of the said proteins are depicted. Finally, Figure \ref{fig:multiview_networks}C shows how these two views fit together. 

\subsection{Single Cell Integration via Anchors}
\label{sec:anchors}
To construct multi-layer networks of networks, computational alignment of various -omics modalities is required. This alignment relies on anchors, which act along the axes of single cell matrices. Although Argelaguet et al.\cite{argelaguetComputationalPrinciplesChallenges2021} define these anchors for combined scATAC, scRNA and CITE-Seq data, we argue that the framework is also highly relevant for single cell proteomics and scRNA alignment, as demonstrated in recent work by Xu et al. \cite{xuScmFormerIntegratesLargeScale2024}.

In \textit{horizontal integration}, rows (genes) act as anchors to align multiple cell-gene matrices. This approach, synonymous with batch-effect removal, serves to eliminate biological or technical noise between equivalent cell sample batches.

\textit{Vertical integration} occurs when multiple biological modalities are measured for a single cell, anchoring matrices along the columns. This is only possible with unambiguous cell identity maintenance across data types, facilitated by combined capture methods. Vertical integration can be further classified into local analyses, detecting individual cross-layer interactions, and global analyses, identifying consensus cross-layer interactions over a wider range of samples.

\textit{Diagonal integration} becomes necessary when no anchors are present in high dimensional space. It poses greater challenges in interpretation and validation but is crucial for integrating single cell experiments where modalities are not simultaneously captured. This strategy relies on the existence of a shared lower-dimensional manifold between modalities, allowing a shared latent space to be found (See \ref{sec:joint_embedding}).

\textit{Mosaic integration} also addresses current technical limitations in capturing different biological data layers from the same cell. In this approach, individual data modalities are profiled from different cell populations within the same biological sample. Missing matrices for respective subsamples are imputed, completing the mosaic. This approach allows for the application of vertical, horizontal, and diagonal integration methods\cite{argelaguetComputationalPrinciplesChallenges2021}.

\subsection{Connecting Layers via Joint Embedding}
\label{sec:joint_embedding}
Diagonal integration, which is currently crucial for large scale multi-omics datasets, relies on joint embedding as a central operation. Also termed joint dimensionality reduction, it effectively addresses the challenge of integrating heterogeneous data types by transforming them into a shared, interpretable latent vector space \cite{leeHeterogeneousMultiLayeredNetwork2020}. Embedding is a high-level classification of methods that includes multiple frameworks outlined in section \ref{sec:integration}. When used specifically in a network-based context, embedding methods include matrix factorisation, network propagation and graph neural network sub-variants. It is critical not only for simplifying the complex structure but also for uncovering relationships that are not readily apparent in the original high-dimensional space. By doing so, joint embeddings facilitate the integration of diverse biological networks, allowing for a more comprehensive understanding of cell behaviour by highlighting how different molecular layers interact\cite{hetzelGraphRepresentationLearning2021}.

\section{Network Inference and Integration Methods\linebreak}
This chapter aims to introduce the main categories of inference and integration methods. The two most relevant method classes in the context of network-based single cell multi-omics are as follows: 
\linebreak
\begin{itemize}
\item \textbf{Methods that infer network structure directly: useful for constructing networks from single cell expression data, corresponding to the unimodal layers.}
\item \textbf{Methods that connect layers: infer missing intra- and cross-layer links. This type of task can be construed as integrating the multilayer network.\linebreak}
\end{itemize}
In line with this, the forthcoming section is split into network\textit{ inference} and network\textit{ integration} subsections. This split, while not without exceptions, generally corresponds to the split between single cell and multi-omics methods, as summarised in \textbf{Table \ref{tab:methods_table}}.

\subsection{Network Inference}
\subsubsection{Knowledge-based Network Inference}\label{sec: kg}

When determining an optimal path towards the solution of a problem, the use of prior knowledge serves as a good starting point. In this sense, knowledge graphs (KG) are invaluable frameworks for capturing complex relational data. They serve as structured graphical representations of heterogeneous graphs, modelling specific relationships between entities. For instance, they can describe the empirically obtained regulatory relationships between genes in gene regulatory networks (GRNs). A fundamental example of KG is Gene Ontology (GO)\cite{thegeneontologyconsortiumGeneOntologyKnowledgebase2023}. In this context, ontology is used to describe representations of known functions and relationships. Precision Medicine Knowledge Graph (PrimeKG), which integrates 20 modalities not limited to molecules, is a versatile example of ontology based KG platform\cite{chandakBuildingKnowledgeGraph2023}.

\begin{table*}[t]
\renewcommand{\arraystretch}{1.7}
\caption{Summary of Methods}
\label{tab:methods_table}
\centering
\small
\begin{tabular}{|p{2cm}|p{3cm}|p{2cm}|p{1.5cm}|p{4.5cm}|p{0.7cm}|}
\hline
\textbf{\raggedright Category} & \textbf{\raggedright Advantage / Disadvantage} & \textbf{Method} & \textbf{Subcategory} & \textbf{Data} & \textbf{Ref.} \\
\hline
\multicolumn{6}{|c|}{\textbf{Network Inference}} \\
\hline
\raggedright \hyperref[sec: kg]{Knowledge Graph-Based} & \raggedright Validated, simple. Static, limited to prior knowledge. & PrimeKG & Multiple KG & Proteins, diseases, pathways, drugs etc. & \cite{chandakBuildingKnowledgeGraph2023} \\
\hline
\multirow{3}{*}{\parbox{2cm}{\raggedright \hyperref[sec:pgms]{Probabilistic Graphical Models}}} & \multirow{3}{*}{\parbox{3cm}{\raggedright Direct interactions, Sparse. Data-dependent.}} & AhGlasso & GGM & Protein & \cite{zhuangAugmentedHighDimensionalGraphical2022} \\
\cline{3-6}
& & piMGM & MGM & RNA, Cancer Subtype, CNV & \cite{manatakisPiMGMIncorporatingMultisource2018} \\
\cline{3-6}
& & BDMCMC & GGM, Bayesian & Gene & \cite{mohammadiBayesianStructureLearning2015} \\
\hline
\multirow{2}{*}{\raggedright \hyperref[sec:boolean]{Boolean}} & \multirow{2}{*}{\parbox{3cm}{\raggedright Dynamic, Executable. Overly Simplistic.}} & PLBIN & Boolean & Protein, RNA, scRNA & \cite{yeInferencingBulkTumor2022} \\
\cline{3-6}
& & mBONITA & Boolean & Protein, Phosphoprotein & \cite{palshikarExecutableNetworkModels2023} \\
\hline
\multicolumn{6}{|c|}{\textbf{Network Integration and Analysis}} \\
\hline
\multirow{3}{*}{\parbox{2cm}{\raggedright \hyperref[sec:mf]{Matrix Factorisation}}} & \multirow{3}{*}{\parbox{3cm}{\raggedright Straightforward, interpretable. Limited to linear transformations.}} & MDN-NMTF & NMF (Tri-factorisation) & miRNA, Disease descriptors & \cite{pengPredictingMiRNADiseaseAssociation2021} \\
\cline{3-6}

\cline{3-6}
& & UINMF & NMF & scRNA, scATAC, Targeted Spatial RNA & \cite{kriebelUINMFPerformsMosaic2022} \\
\hline
\multirow{3}{*}{\parbox{2cm}{\raggedright \hyperref[sec:netprop]{Network Propagation}}} & \multirow{3}{*}{\parbox{3cm}{\raggedright Long-range interactions. Topology-dependent and expensive}} & BRWRMHMDA & Biased RWR & miRNA, Disease descriptors & \cite{quBiasedRandomWalk2021} \\
\cline{3-6}
& & \textit{unnamed} & RWR, HD & RNA, Protein & \cite{cowenNetworkPropagationUniversal2017} \\
\cline{3-6}
& & HotNet2 & HD & Gene & \cite{leisersonPancancerNetworkAnalysis2015} \\
\hline
\multirow{2}{*}{\parbox{2cm}{\raggedright \hyperref[sec:gnn]{Graph Neural Networks}}} & \multirow{2}{*}{\parbox{3cm}{\raggedright Flexible, applicable to large datasets. Limited interpretability.}} & DeepMAPS & HGT & scRNA, scATAC, CITE-Seq & \cite{maBiologicalNetworkInference2021} \\
\cline{3-6}
& & GLUE & GVAE & scRNA, scATAC, snmC & \cite{caoMultiomicsIntegrationRegulatory2021} \\
\hline
\multirow{2}{*}{\raggedright \hyperref[sec:combos]{Combination}} & \multirow{2}{*}{\parbox{3cm}{\raggedright Synergies, complicated architectures}} & HGNNLDA & RWR, HGT & lncRNA, Disease descriptors & \cite{hongshiHeterogeneousGraphNeural2022} \\
\cline{3-6}
& & HGAN & HD, GAT & Drug-Target Interactions & \cite{liHeterogeneousGraphAttention2022} \\
\hline
\end{tabular}
\end{table*}

It includes ontologies for diseases, drugs, phenotypes, proteins and biological pathways, and clinical data. This serves as an effort to unify the scattered knowledge bases contained in non-standardised repositories and publications. The resulting knowledge graph, which consists of approximately 129,000 nodes and over 4,000,000 edges, allows the inference of associations between various diseases and drugs. While KGs generally implement automation for large scale graph generation, the underlying input is still reliant on manual curation.

\subsubsection{Probabilistic Graphical Models}\label{sec:pgms}
In contrast to knowledge-driven graph generation, a fundamentally data-driven approach of graph generation  comes in the form of relevance networks\cite{werhliComparativeEvaluationReverse2006}\cite{yeInferencingBulkTumor2022}. Such correlation-based methods infer interactions from statistical relationships between entities. In the case of RNA data, co-expression of genes could act as correlation measures. Most commonly, Pearson correlation or regression is used to determine the proximity of pairs of genes in cell-space (See section \ref{sec:building_sc_from_matrices}). 

Correlation analysis generates many spurious connections, since many pair dependencies are conditioned on the presence of a third entity. Therefore, partial correlation-based methods deduce the conditional dependencies, or partial correlations, between entities[4]. This effectively prunes many connections, resulting in a sparser network that more accurately reflects direct relationships, such as physical interactions or participation in shared biological pathways. Partial correlation-based models are an instance of a broader class of flexible methods, namely probabilistic graphical models (PGMs)\cite{agamahComputationalApproachesNetworkbased2022}\cite{huynh-thuGeneRegulatoryNetwork2019}\cite{saint-antoineNetworkInferenceSystems2020}, which include both undirected and directed (causal) models. In fact, the utilisation of partial correlations, or conditional dependencies, serves as the basis for graphical models \cite{lauritzenGraphicalModels1996}\cite{haweInferringInteractionNetworks2019}\cite{altenbuchingerGaussianMixedGraphical2020}. 

If the variables are assumed to be normally distributed, we arrive at a form of PGM known as Gaussian Graphical Models (GGM). Here, connections between nodes are simply inferred via an inverse covariance matrix $\Sigma^{-1}$. Recent extensions to GGMs, such as the Augmented High-Dimensional Graphical Lasso (AhGlasso), have been developed to incorporate prior knowledge (knowledge graphs) of protein-protein interactions with edge weights for improved global network learning in high-dimensional settings \cite{zhuangAugmentedHighDimensionalGraphical2022}. The Gaussianity constraint is relaxed in Mixed Graphical Models (MGM)\cite{lauritzenGraphicalModels1996}, which are a subset of PGMs that allow independent distribution modelling for different data types such as discrete and categorical data\cite{manatakisPiMGMIncorporatingMultisource2018}. The popularity of these methods stems from their straightforward interpretation via conditional dependencies, allowing the distinction between direct and indirect effects, as well as their computational efficiency\cite{altenbuchingerGaussianMixedGraphical2020}. Variations on the method of calculating correlations have been employed in literature to arrive at new forms of PGMs.

\subsubsection{Temporal Data-based Edge Directionality Inference}
PGMs are straightforward and comparatively efficient in terms of computation. However, inferring directionality can only be done with either time-stamped data or prior knowledge, as seen in the above examples. If such information is available, it can be leveraged by causal models. The networks constructed using causal models have the benefit of allowing for directed relationships between nodes\cite{hetzelGraphRepresentationLearning2021}, as well as the detection of subnetworks\cite{agamahComputationalApproachesNetworkbased2022}. To infer the direction from temporal information, as demonstrated in models such as SINCERITIES\cite{papiligaoSINCERITIESInferringGene2018}, a time-stamped omics dataset is required. In this approach, regression analysis, a statistical method to estimate the dependency relationships among variables, is employed. Specifically, ridge regression—a regularised form of linear regression less prone to overfitting—is used. This form of regression adds a penalty to the optimisation process, which results in reduced complexity of the model and prevents fitting noise in the data. Subsequently, partial correlation is used to differentiate between upregulation and downregulation of genes.

\subsubsection{Knowledge-based edge directionality inference}
In the absence of temporal information, prior knowledge could be used to establish directionality between nodes. One such correlation-based method focusing on single cell transcriptomics was developed by Iacono et al. \cite{iaconoSinglecellTranscriptomicsUnveils2019}. This method employs a specialised form of correlation based on the transformed variables (Z-scores). A knowledge graph approach informs directionality using gene ontology, providing a subset of "regulators of gene expression". This method is intended for global, large scale network inference and was tested on Alzheimer's and diabetes datasets. Prior knowledge can also inform Bayesian belief models. Underlying these is the notion of conditional probability. Not to be confused with conditional dependence of GGMs, the conditional probability P (Y |X) of two genes X and Y corresponds to a directionality of X → Y . This type of directed relationship expresses the probability of gene Y being expressed, given the knowledge that X is being expressed.  Directed edges can then be established between a set of multiple genes, resulting in a Bayesian network, or directed acyclic graph (DAG)\cite{saint-antoineNetworkInferenceSystems2020}. An example of this is seen in work by Mohammadi et al.\cite{mohammadiBayesianStructureLearning2015}, where Bayesian node-wise selection (BANS) exploits prior knowledge for directed inference. This Bayesian strategy is used on top of a multi-layer GGM, where BANS is used to estimate directed and undirected edges in the network.
A major drawback of these Bayesian methods is their computational cost, which limits them to the analysis of small datasets. Furthermore, cyclical relationships in the form of feedback loops cannot be captured with most methods, since the framework utilises DAGs\cite{yeInferencingBulkTumor2022}. This limitation detracts from the applicability for GRNs, which are characterised by such cyclical motifs. 

\subsubsection{Boolean Control Networks}\label{sec:boolean}
Boolean control networks (BCNs)\cite{schwabConceptsBooleanNetwork2020} are a class of models that address the issue of scaling, while at the same time offering unique benefits over correlation models in terms of modelling capability. In BCNs, Gene activation states are binarised from expression signatures, upon which logical update functions are found to inform gene relationships\cite{yeInferencingBulkTumor2022}. The simplifying nature of binarising node states facilitates efficient scaling of network size \cite{gyoriWordModelsExecutable}. Belonging to the class of executable models, BCNs also allow for dynamic simulation of information flows. Recently, the inference of Boolean networks has been streamlined, making the process less computationally expensive \cite{parulmaheshwariInferenceBooleanNetwork2022}. Ye and Guo \cite{yeInferencingBulkTumor2022} developed a Prediction Logic Boolean Implication Network (PLBIN), which improves upon basic models by adding implication relationships between variables via logical rules. Such implication relationships can be expressed as "A high level of expression for gene A implies the same for gene B", inferring their relationship. The use of a contingency table, constructed from the expression values of two genes, aids in selecting optimal implication functions by calculating their respective precision. Therefore, links between each gene pair are inferred from the contingency table of said pair.

In the field pertaining to biological research, Palshikar et al. \cite{palshikarExecutableNetworkModels2023} introduced a Boolean network-based method for multi-omic integration and relation modelling. Multiomics Boolean Omics Network Invariant-Time Analysis, or mBONITA, incorporates knowledge graphs for topology-based pathway analysis. It captures transcriptomics, proteomics and phosphoproteomics data. As an executable model, mBONITA enables network perturbation analysis in a multi-omics setting. In this sense, network perturbation allows for the assessment of dynamic response to changes in gene and pathway expression. While Boolean networks offer benefits in terms of executability, binarising gene expression leads to loss of information, since real expression values are continuous.

\subsection{Network Integration and Analysis}\label{sec:integration}
So far, we have touched upon the methods that allow the construction of a graph using -omics data input and prior knowledge associated with the data. In this section, we will focus on methods that take multiple of these inferred structures, each associated with a different -omics class, and integrate them into a combined  multilayer graph. 

\subsubsection{Matrix Factorisation}\label{sec:mf}
Joint embedding is central to heterogeneous data integration. Genes may be described as points in cell-space, where the dimensionality of the space is determined by the number of measurements (See section \ref{sec:building_sc_from_matrices}). The embedding operation enables genes to be cast into a shared space defined by measurements on multiple levels. For example, a gene's expression may be measured via scRNA-Seq, as well as single cell proteomics, resulting in two separate data matrices. The integration of these disparate data types commonly involves dimensionality reduction, which simplifies the feature space by assuming that phenomena of interest are described by a core set of latent features. A key technique in this process is matrix factorization (MF), a broad category of methods that reduce a matrix into lower-dimensional matrices via linear transformations. Matrix factorisation describes each cell as a product between a vector of its -omics values (proteomics, RNA, etc.) and a vector of  the cell's "core" features shared across data types. Fig. \ref{fig: 3intmethods}A visualises this factorisation, where the multiplication on the right hand side shows a common factor and the individual -omics matrices. 

Non-negative matrix factorization (NMF), which imposes a positivity constraint on the factors, is widely used in biological network analysis for tasks like multi-omic data integration or module detection (the latter of which is used to probe the modular structure of gene networks that supports biological functionality). 

One of the downstream tasks of such data integration is network inference. A common problem for relation inference is the identification of associations between molecular factors and disease states. This form of heterogeneous link prediction has been demonstrated by Peng et al. \cite{pengPredictingMiRNADiseaseAssociation2021}, making use of non-negative matrix tri-factorisation for the prediction of miRNA - disease associations. Their method constructs a heterogeneous network with miRNA and disease information, based on known associations between diseases and known relation of miRNA to said diseases. Furthermore,  latent vector representations for both are learned. By multiplying these vectors, the association probability is calculated. 

Recent research has employed matrix factorization in a single cell multi-omics context where vertical integration is not possible due to batch discrepancies. Introduced as the "mosaic integration" problem by Argelaguet et al. (section \ref{sec:anchors}), this form of integration targets data from separate experiment batches performed across different biological levels. To address the challenges resulting from input matrices with discrepant dimensions, Kriebel et al. \cite{kriebelUINMFPerformsMosaic2022} developed a non-negative matrix factorisation method to integrate single cell multi-omic datasets. The key contribution of the method, termed UINMF, is to utilise an unshared metagene matrix. This unshared matrix allows the incorporation of features that are only present in a subset of data types, whereas previous methods could only use data that was shared by all modalities. UINMF allows for the integration of single cell RNA-Seq, spatial transcriptomic, single-nucleus chromatin accessibility and mRNA expression sequencing (SNARE-Seq) \cite{chenHighthroughputSequencingTranscriptome2019} and cross-species datasets. 

Thus, non-negative matrix factorisation methods are effective and straightforward, offering interpretability due to the additivity of its factor matrices \cite{leeLearningPartsObjects1999}. However, a major limitation of current matrix factorisation methods is the assumption of linearity. The decomposition of the data matrices into their latent factors is linear, which means that the linear reconstruction of said matrices cannot capture non-linear, complex relationships across layers \cite{leeHeterogeneousMultiLayeredNetwork2020}. Furthermore, exceedingly large matrices are prohibitive, limiting the use of this framework for genome-scale multi-omics data at a single cell level.

\subsubsection{Network Propagation}\label{sec:netprop}
\begin{figure*}[t]
    \centering
    \includegraphics[width=0.8\textwidth]{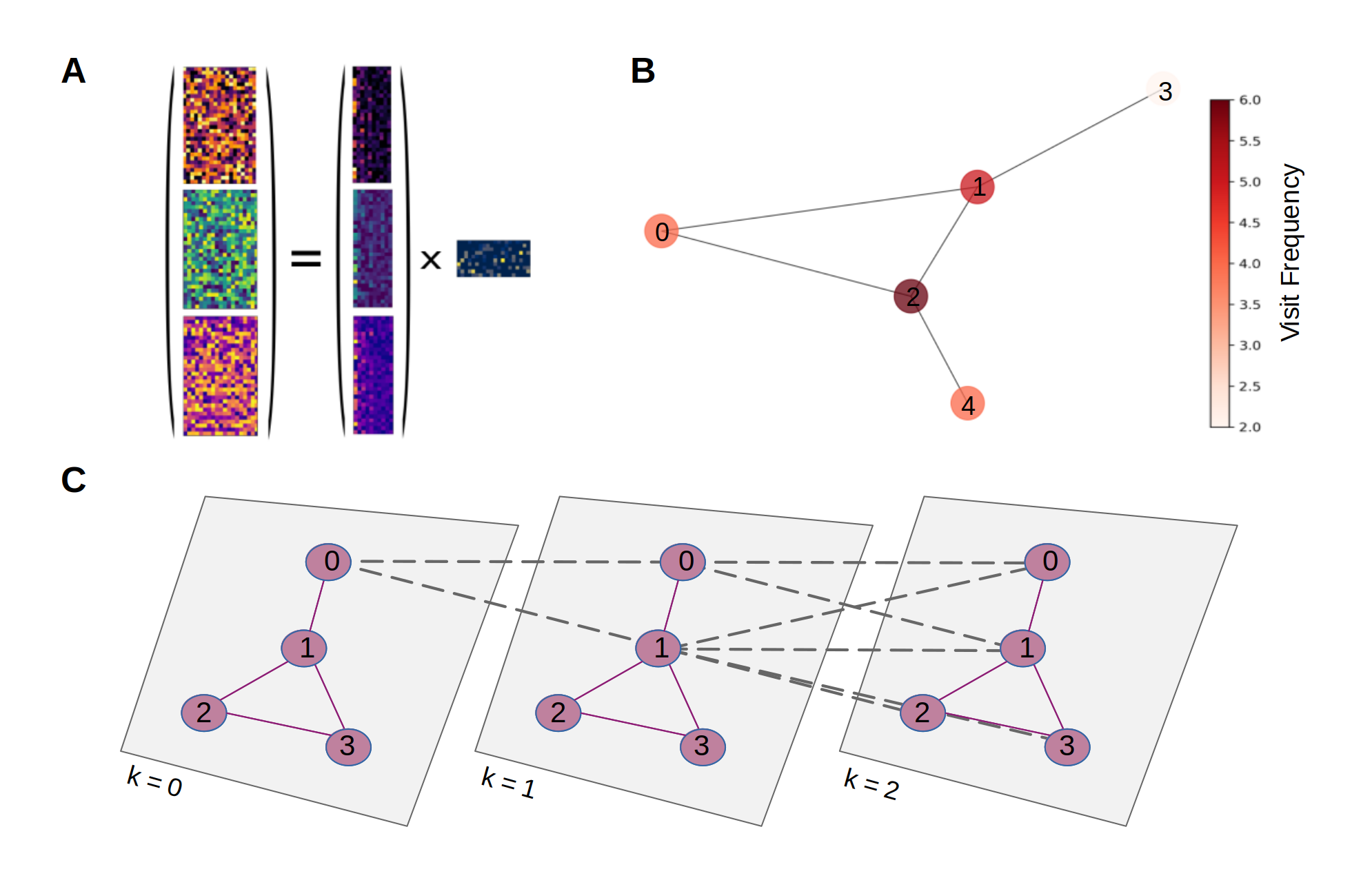}
    \caption[Comparison of Integration Methods]{Comparison of Integration Methods. \textbf{(A)} The decomposition of 3 data matrices from different sources into lower dimensional matrices via factors. \textbf{(B)} Topological node information obtained via random walks, where colour intensity indicates visit frequency. \textbf{(C)} 3 layers of a graph neural network, indicated by \emph{k}. Dashed lines indicate message passing, where information is aggregated at node 0.}
    \label{fig: 3intmethods}
\end{figure*}

An alternative class of methods, summarised as network propagation, can address the challenge of capturing non-linear relationships. Emphasised as a powerful and flexible framework that has found use across disciplines under various guises, including PageRank and heat diffusion (HD), network propagation is a dominant technique in systems biology \cite{hetzelGraphRepresentationLearning2021}. An intuitive instance of this framework is the random walk (RW). Here, a "walker" moves from node to node in the network, its path being indicative of the nodes' connectivity and position in the network. Figure \ref{fig: 3intmethods}B depicts the averaged results of a random walk starting at such a source node (Node 0). Highly connected nodes will be visited frequently, indicating their status as hubs. Random walk with restart (RWR) is a variation on this approach which adds a fixed probability of returning to the source node at each step, thereby confining diffusion to local neighborhoods even at steady state.

Topological information obtained from these algorithms can be used to infer missing links in the network, making it a critical tool for multilayer network inference \cite{zhouPredictingMissingLinks2009}. By propagating information across layers and long distances, global interactions are revealed. Rather than investigating pairwise interactions between nodes, network propagation is capable of detecting complex patterns involving multiple nodes, such as functional modules that drive disease. 

Cowen et al. \cite{cowenNetworkPropagationUniversal2017}elucidate the efficacy of this method for detecting cancer driver genes, whose somatic mutations contribute to tumour development.  Underlying this approach is the notion that genes related to a disease are more likely to interact with each other than with randomly selected genes. Therefore, biological prior knowledge, such as known driver genes of cancer, is superimposed on the network. Using these nodes as sources, the information is propagated across the network, amplifying the signal and revealing clusters of highly significant nodes.

Recently, Charmpi et al. \cite{charmpiOptimizingNetworkPropagation2021} have sought to optimise network propagation for multi-omics integration. By framing it as a network normalisation method, the authors highlight it as an approach to parameter selection, integrating both proteome and transcriptome data. They demonstrate the efficacy and robustness of the approach for investigating prostate cancer progression. 

The use of heterogeneous network propagation is also central to BRANEnet \cite{jagtapBRANEnetEmbeddingMultilayer2022}, which utilises several methods presented in this review to obtain joint embeddings for multi-omics data. The model efficiently captures high-dimensional, complex multi-omics data patterns, enabling integrated omics network inference, module detection and disease-gene associations. The authors test it on both transcriptomics and targeted metabolomics data. The counterpoint to its flexibility comes due to the algorithm suffering from high computational cost for larger network structures, rendering it unfeasible for certain tasks on genome-scale heterogeneous networks \cite{leeHeterogeneousMultiLayeredNetwork2020}. Furthermore, the algorithm is highly dependent on the quality of the inferred network, as it is sensitive to differences in topology.

\subsubsection{Graph Neural Networks}\label{sec:gnn}
GNNs and their sub-variants, graph convolutional nets (GCN) and graph attention nets (GAT), form a powerful class of methods for capturing complex interactions and nonlinearities in biological data\cite{zhangGraphNeuralNetworks2021} (See Fig. \ref{fig: 3intmethods}C for a schematic depiction of spatial graph convolution, or message passing). GNNs are flexible tools that have proven to be capable of handling large-scale, heterogeneous and complex networks. They are able to extract the topology and key features of data at a massive scale \cite{huHeterogeneousGraphTransformer2020}. While typically operating on non-Euclidean data, GCNs in the form of convolutional neural networks demonstrate effective extraction of high-level information from Euclidean data in the form of images \cite{lecunGradientBasedLearningApplied1998}. GNNs have been employed for single cell transcriptomic data, inferring cell-cell communication, cell differentiation and disease state prediction\cite{maBiologicalNetworkInference2021} \cite{yinScGraphGraphNeural2022}\cite{xingMultilevelAttentionGraph2022}. 

Regarding heterogeneous network inference, a method was developed by Shi et al. \cite{hongshiHeterogeneousGraphNeural2022} for constructing a network of lncRNA and miRNA and their associated diseases. By learning node embeddings in a shared latent space, relations between the different biological molecules can be predicted via similarity metrics in this space. Cao and Gao recently introduced graph-linked unified embedding (GLUE) \cite{caoMultiomicsIntegrationRegulatory2021}, which is capable of diagonal data integration guided by knowledge graphs (see section \ref{sec:building_sc_from_matrices} for background on diagonal integration). Taking single cell multi-omics data as input, cell states are modelled as low-dimensional cell embeddings. These embeddings share a common latent space across modalities. Biological prior knowledge is used to incorporate regulatory interactions in the form of a guiding knowledge graph. A graph variational autoencoder (GVAE) is then used to learn feature embeddings from these cross-layer relationships. Graph VAEs are deep neural networks trained specifically to embed graphs from a higher dimensional vector space into a lower dimensional one, simplifying their representation\cite{kipf_variational_2016}.  Finally, -omics data is reconstructed from the inner product of the embedded cell and feature matrices. The strategy posed by GLUE and similar methods come as a response to the ever-increasing volume of data from various unaligned sources. While it efficiently integrates triple-omics sources, it simultaneously suffers from the data dependency that marks neural network-based methods. Despite this, GLUE exhibits remarkable performance and robustness to hyperparameter settings, as evaluated on benchmark datasets. Hyperparameter settings included a range of embedding dimensionalities, as well as hidden layer depths. The latter refers to the number of layers used in the autoencoder.     

GATs, in combination with meta-paths, particularly hold promise for heterogeneous data analysis in the form of heterogeneous graph transformers (HGT) \cite{huHeterogeneousGraphTransformer2020}. This architecture is employed in the context of single cell multi-omic networks through DeepMAPS \cite{maBiologicalNetworkInference2021}. Explicitly modelling topological information shared between modalities, the model constructs a heterogeneous network of cells and genes from multimodal data (Single Cell RNA-Seq, CITE-Seq and ATAC-Seq). Following network embedding, attention scores are calculated to determine the importance of specific cell-gene relationships. From this, cell-specific gene networks are inferred, allowing downstream clustering and regulatory network inference \cite{stefanstanojevicComputationalMethodsSinglecell2022}. Therefore, the attention mechanism serves as a selector for assigning cell identities to gene networks. This provides interpretability of the model, enabling the identification of nodes (genes) most relevant for defining cell-state differences. The method was employed to infer cell-type specific networks in lung tumour leukocyte data, as well as small lymphocytic lymphoma.

\subsection{Integration Method Combinations}\label{sec:combos}
Exemplifying a combination of GATs and network propagation, Multi-Hop Attention GNN (MAGNA) \cite{wangMultihopAttentionGraph2021} diffuses attention scores across a knowledge graph. This results in a diffusion prior on attention values and increases the receptive field across the network. MAGNA has been adapted for heterogeneous biological networks by Li et al. \cite{hongshiHeterogeneousGraphNeural2022}, whose model aims to predict drug-target associations by diffusing attention within and across network layers, generating a prior for the GAT, which in turn performs link prediction, node classification, or other tasks. A specific use case for GNNs comes in the form of handling noisy data, which is a frequent issue for protein-protein interaction networks. For this purpose, a graph variational autoencoder was developed by Yao et al. \cite{yaoDenoisingProteinProteinInteraction2020} for denoising such datasets and improving network inference accuracy.

Despite their flexibility and power, it must be noted that GNNs are still vulnerable to the imbalances in biological data. As data-driven models, they suffer from undertraining on small datasets. Furthermore, while GATs have proven to allow for higher interpretability, GNNs in general still suffer from the black box nature of machine learning \cite{zhangGraphNeuralNetworks2021}. Additionally, traditional GNNs suffer from certain limitations such as oversmoothing, oversquashing and bottlenecks \cite{alonBottleneckGraphNeural2021}, where aggregation over long distances results in information loss. Oversmoothing refers to the convergence of node values towards a single point with successive iterations, limiting the depth of GNNs. Oversquashing occurs when distant nodes fail to communicate with each other effectively due to a limited amount of information being incorporated into the fixed-size feature vectors of nodes. The solution to these issues has been explored by the authors of GRAND \cite{chamberlainGRANDGraphNeural2021}, who developed a hybrid model incorporating diffusion into the GNN architecture. By controlling the smoothing process of information propagating across the graph via message passing, the convergence of node values and loss of heterogeneity is avoided. The use of advanced solvers for diffusion partial differential equations (PDEs) also improves the stability of the algorithm. These solvers, such as implicit schemes and adaptive step size methods like Runge-Kutta, provide unconditional numerical stability for any step size and can automatically adjust their parameters to maintain accuracy while minimising computational cost. Oversquashing is mitigated via graph rewiring, which improves information flow and combats its loss over distances. While this model has not yet been used in molecular biological settings, it demonstrates the advantages of creating hybrid models that incorporate components from multiple of the aforementioned classes of tools.   

\section{Method Validation\linebreak}
While all of the above-listed methods will provide some output, it is essential to ensure that these results align with biological reality to a sufficient degree, both in inference and integration settings. A review by Chen et al. \cite{chenEvaluatingMethodsInferring2018} revealed that the majority of bulk inference tools are no longer adequate in the single cell era. Therefore it is necessary to develop inference tools which take the necessary considerations, such as zero-inflation, into account. 

One way to evaluate performance is via comparison to ground truth networks. Chen et al. utilised synthetic networks generated with GeneWeaver \cite{bakerGeneWeaverWebbasedSystem2012} and gene relation networks from publicly available experimental data. Corresponding datasets describing both of these network types were generated from the network structures. This allowed for the true network structure to be known, as well as providing input data for the models to be tested. Model performance was evaluated using Receiver Operating Characteristic (ROC) and Precision-Recall (PR) curves. Results showed that performance on experimental data was poor across the board, while single cell-specific methods performed marginally better on one of the simulated datasets. 

A similar approach for benchmarking was employed by Pratapa et al. \cite{pratapaBenchmarkingAlgorithmsGene2020a} via comparison with ground truth networks taken from synthetic networks, literature-curated Boolean models and diverse transcriptional regulatory networks. The authors developed BEELINE, a comprehensive evaluation framework to assess GRN reconstruction performance for single cell expression data. The framework allows comparison of accuracy, stability and efficiency of methods across single cell datasets. By providing a uniform pipeline that includes pre-processing, parameter estimation and post-processing, comparability is ensured. The main performance metric is the Area Under Precision-Recall Curve (AUPRC) ratio, which is the AUPRC of the method to be tested divided by that of a random predictor. The ratio ranged from 1.3 to 4.8 for the best-performing methods, which include SINCERITIES, SINGE and SCRIBE\cite{papiligaoSINCERITIESInferringGene2018}\cite{deshpande_network_2022}\cite{qiu_inferring_2020}.

Another way to validate inferred networks is by comparing the network overlap between two independent datasets originating from the same biological condition. Kang et al. \cite{kangEvaluatingReproducibilitySingleCell2021a} highlight the importance of this approach by going beyond simulated datasets for validation via a reproducibility study. The rationale in this study is that the inferred networks should strongly overlap with each other, given the shared biological ground truth. Cell populations included human retina and T cells from colorectal cancer (CRC). Six methods were evaluated, among which GENIE3 \cite{huynh-thuInferringRegulatoryNetworks2010} and GRNBoost2 \cite{moermanGRNBoost2ArboretoEfficient2019a} showed the highest overlap between the datasets and therefore higher reproducibility.	 Specifically, for the human retina, GENIE3 achieved a 45.9\

Validating multi-layer network integration methods may be done in several ways. For example, the authors of BRANEnet\cite{jagtapBRANEnetEmbeddingMultilayer2022} tested the performance of predicting TF-target interactions between the protein and RNA layers. To generate test sets, 50\

Wen et al. \cite{wenGraphNeuralNetworks2022} assessed the ability of GNNs in performing joint embedding via six metrics, split into two classes. Biology conservation metrics evaluate how well the methods retain biological information, while batch removal metrics measure the ability to remove batch effects and integrate data types. For an example of the former, retention of biological information can be assessed via clustering cells in embedded space and comparing the clusters with known cell labels. In an example of evaluating NMF performance for sample clustering, Cantini et al. \cite{cantiniBenchmarkingJointMultiomics2021} examined the factor matrix containing biological samples in latent space. Clusters inferred in latent space were compared to ground-truth labels using the Jaccard Index (JI) and Adjusted Rand Index (ARI). Their results highlight the performance of IntNMF, which achieved near-perfect clustering performance with JI scores of approximately 1 and ARI scores between 0.9-1.0, outperforming other dimensionality reduction methods for clustering tasks. However, IntNMF was less effective at other analyses - for clinical annotation tasks it ranked among the top three methods in only 2 out of 10 cancer types, and for biological pathway analysis it achieved top three performance in only 3 out of 10 cases.

\section{Outlook\linebreak}
\label{M_end}
The challenge of heterogeneous network inference is essentially a data integration problem. While the utility of network propagation methods like heat diffusion or random walks has been well-established in the field, recent developments in graph neural networks (GNNs) have also shown promise. Particularly, GNNs such as Graph Attention Networks (GATs) have demonstrated their flexibility in handling heterogeneous data, crucial for tasks like cross-layer link prediction. 
While network propagation and graph neural nets are often presented as independent frameworks in the computational biology literature, they are in fact fundamentally related and mutually beneficial. Implementing diffusion into GNNs overcomes detrimental shortcomings and allows distant nodes to communicate more effectively. This is especially relevant for multi-layer relation inference, so that nodes across layers effectively propagate information regardless of their distance. Single cell multi-omics methods remain underrepresented in the literature, mainly due to the limitations and recency of combined capture methodologies. However, as datasets of this category are continuously generated, it is crucial for computational methods to remain up to speed. Coupled single cell sequencing for multiple biomolecule types allows the creation of multi-scale networks in the form of molecular networks within cell-cell interaction networks. These can be modelled for multiple layers, capturing essential biological dependencies between molecule types. 

An extensive list of modelling frameworks that can be applied to either of these tasks has been presented in this review. This includes data-driven, probabilistic and causal models in the graph inference category, as well as embedding-focused tools such as matrix factorisation, diffusion and GNNs in the graph integration category. 
However, before a truly representative mechanistic model can be obtained, several hurdles must be overcome. Inferring directionality of interactions is difficult and goes beyond the simple inference of interactions. The two-pronged nature of multi-scale, multi-layer complexity further complicates modelling choices. Cell-cell interactions cannot be simultaneously inferred with gene-protein-metabolite interactions at this point in time, since approaches are divergent and are highly sophisticated in their own regard. To tackle the challenges that arise from such an increase in complexity, a combination of tools is necessary. Beyond picking the right tools from an assorted toolkit, the tools must mutually enhance each other to overcome their respective limitations and permit the accurate modelling of biological phenomena of interest.  

\end{multicols}

\bibliographystyle{unsrt}
\bibliography{My_Library}  

\end{document}